\documentclass{article}
\usepackage{spconf,amsmath,graphicx,booktabs,url}

\title{\textbf{Which Layer to Trust? Layer Selection in Frozen Speech Encoders\\
for Singing-Voice Deepfake Detection under Signal Corruption}}

\name{Benjamin Hurt$^{*}$, Yefei Yang$^{*}$\thanks{$^{*}$Equal contribution (co-first authors).}}
\address{University of California, Santa Barbara}

\begin{document}
\ninept
\maketitle

\begin{abstract}
Frozen self-supervised speech encoders have been frequently used for audio deepfake detection. However, the layers are usually chosen from a development set. We examine this choice for singing-voice deepfake detection on CtrSVDD. We train one detection head per layer, plus a learned layer mix, on frozen XEUS (3--4 seeds) and XLS-R (2 seeds). We test on clean audio and under noise, reverberation and lossy coding. On clean audio, the middle layers beat the last layer in both encoders and the learned mix does not beat the best single layer. The layer chosen on the development set is 2.3 (XEUS) and 4.1 (XLS-R) percentage points of EER worse than the best test layer. Under noise and coding, XEUS's best layers stay in the middle, while under reverberation its upper layers catch up, whereas XLS-R's best layer moves shallower. In an exploratory analysis, XEUS detects a commercial synthesis attack best with an earlier layer, especially under noise. The layer therefore cannot be chosen reliably on an in-domain development set and clean audio alone.
\end{abstract}


\section{Introduction}
\label{sec:intro}

Self-supervised learning (SSL) encoders have been used as a standard front end for audio deepfake detection~\cite{wang2022ssl,tak2022wav2vec}. When the encoder is frozen, a light back end is trained on its hidden states. Frozen XEUS with a multi-head factorized attentive pooling (MHFA) back end, for example, has been reported to generalize across speech deepfake benchmarks at a small fraction of the trainable parameters of fine-tuned systems~\cite{ism2026}. Such a system still has to decide which hidden layer, or which mixture of layers, to use. Layer-wise analyses show different depths of speech SSL models encode different information~\cite{pasad2021}, and a recent study of SSL models for audio deepfake detection found lower layers to be the most discriminative across languages and deepfake types, including songs~\cite{elkheir2025layer}. Benchmarks often sidestep the choice with a learned weighted sum over layers~\cite{superb2021}.

Two aspects of this choice are rarely examined together. First, layers are compared on clean test audio, whereas in deployment, a detector receives audio degraded by background noise, rooms and lossy codecs --- channel and codec effects are an established source of degradation in speech anti-spoofing~\cite{asvspoof2021}. Second, the layer, like any hyperparameter, is usually chosen on a development set that contains the training attacks. A layer that does well on those attacks may not generalize to unseen ones.

We study both questions for singing-voice deepfake detection (SVDD), where detectors made for speech do not transfer directly~\cite{singfake2024,ctrsvdd2024} and the strongest challenge systems rely on SSL front ends~\cite{svdd2024}. Using the CtrSVDD benchmark~\cite{ctrsvdd2024}, we train a head on every layer of a frozen encoder and on a learned layer mix, with one recipe and one shared data order per seed, for XEUS~\cite{xeus2024} (3--4 seeds per condition, clean audio and nine corruptions) and XLS-R~\cite{babu2022xlsr} (2 seeds, clean audio and three corruptions). We find that: (i)~in both encoders, a middle layer beats the last layer on clean audio, and a learned layer mix does not beat the best single layer; (ii)~the development set does not reliably pick the layer: it barely separates the middle layers, and its pick costs 2.3 and 4.1\,pp of main EER; (iii)~the best depth shifts with the signal condition, differently in each encoder and most clearly under reverberation; and (iv)~in an exploratory analysis, XEUS's best layer for an out-of-domain commercial attack differs from that for the other attacks, increasingly so under noise.

\section{Experimental Setup}
\label{sec:setup}

\textbf{Data and metrics.} CtrSVDD~\cite{ctrsvdd2024} contains 16\,kHz mono singing clips split into training (84,404), development (43,625) and test (92,769) sets. Training and development share attacks A01--A08; the test set uses A09--A14, none of which occurs in training. A14 is produced with a commercial singing-voice synthesizer and was excluded from the challenge ranking because it is out of domain~\cite{svdd2024}. We report equal error rate (EER; lower is better) in two ways: \emph{main EER}, all 13,596 bona fide clips against the 53,983 A09--A13 spoofs; and \emph{A14 EER}, the same bona fide clips against the 25,190 A14 spoofs. Like the official challenge ranking, main EER excludes A14; unlike it, both our metrics keep the ACESinger bona fide clips, which the ranking excludes~\cite{svdd2024}.
Each clip is truncated or zero-padded to its first 64,600 samples (about 4\,s).

\textbf{Encoders and heads.} XEUS~\cite{xeus2024} is a 19-layer E-Branchformer encoder pretrained on over one million hours of speech in 4,057 languages with masked prediction, denoising and dereverberation objectives. XLS-R 300M~\cite{babu2022xlsr} is a 24-layer wav2vec\,2.0 Transformer pretrained on 436k hours in 128 languages, without denoising or dereverberation objectives. We keep both encoders frozen. Only the heads, fusion modules, and mix weights are trained. We take every layer's output, and in both encoders, the last layer is taken after the final layer normalization.\footnote{ESPnet's XEUS front end standardizes the input waveform over the whole batch, so a clip's score depends on the other clips in its batch. We standardize each clip separately. On earlier checkpoints of this pipeline, this changed EER by 0.15--0.18\,pp (main) and 1.1--1.4\,pp (A14).} Each single-layer head applies a linear projection and layer normalization, multi-head factorized attentive pooling (MHFA)~\cite{peng2022mhfa} ($K{=}64$, $H{=}8$), a second layer normalization and a linear classifier. The learned-mix head pools a softmax-weighted sum of all layers, starting from uniform weights, and is trained from scratch with the same recipe. For XEUS only, we also train the same heads inside the SSL--CQCC cross-attention fusion of~\cite{elkheir2025tvot}, which adds constant-Q cepstral features.

\textbf{Training.} All heads of a seed are trained in one job. Each batch passes through the encoder once and goes to every head. The heads of a seed thus see the same data in the same order, while each keeps its own optimizer, gradient clipping, and early stopping, so differences between layers within a seed are paired. Heads use AdamW (learning rate $10^{-4}$, weight decay $10^{-4}$; $10^{-3}$ without decay for the mix weights), a cosine schedule with two warm-up epochs, at most 20 epochs, batch size 24 and class-weighted cross-entropy. Each head keeps the checkpoint with the lowest development EER and stops once it has not improved for six epochs. We train four seeds for XEUS and two for XLS-R.

\textbf{Test-time corruptions.} Corruptions are applied to test audio only, deterministically per clip and before truncation; training uses clean audio only. Additive noise comes from the MUSAN noise subset~\cite{musan2015} at 20, 10, 5 and 0\,dB SNR, measured over the full clip, so the SNR of the scored 4\,s segment varies. Reverberation convolves each clip with one of 325 real room impulse responses from the RIRS\_NOISES collection~\cite{ko2017} (RWCP, AIR and REVERB 2014) and restores the input level. Lossy coding uses MP3 at 128 and 32\,kbit/s and Opus at 32 and 12\,kbit/s, decoded back to 16\,kHz. For XEUS, all four seeds are scored on clean audio, 5 and 0\,dB noise and reverberation, and three seeds on the other six conditions. XLS-R is scored on clean audio, 5 and 0\,dB noise and reverberation.

\textbf{Analysis.} We report seed means and standard deviations, and compare layers through per-seed paired differences, with the number of seeds that agree on the sign. Differences are in percentage points (pp). To compare encoders of different depth, we use relative depth (layer index divided by the number of layers). The \emph{development-selected layer} has the lowest mean development EER, a rule we fixed in advance. The \emph{best test layer} has the lowest mean test EER; it is chosen in hindsight and serves only as a reference. Because the A09--A13 spoofs come from only 11 of the 51 test singers, we also check whether key differences survive singer resampling: a class-stratified bootstrap over test singers ($B{=}1000$, paired across heads) gives 95\% intervals, and an asterisk in Table~\ref{tab:grid} marks differences whose interval excludes zero. The A14 analysis (Sec.~\ref{ssec:a14}) was not planned: we chose its comparisons after seeing three XEUS seeds, and the fourth seed, trained later, serves as a held-out check. We also choose the best A14 layer on one half of the test clips and test it on the other, in both directions (halves split by clip ID, so singers overlap).

\section{Results}
\label{sec:results}

\begin{figure}[t]
\centering
\includegraphics[width=\columnwidth]{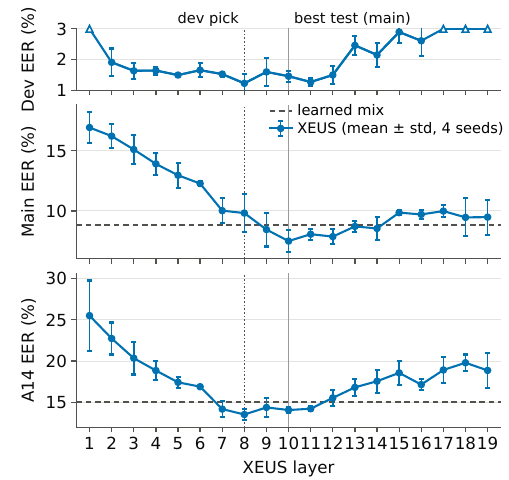}
\caption{XEUS-only heads, clean test set: development, main and A14 EER per layer (mean $\pm$ std over four seeds). Dashed: learned layer mix. Vertical lines: development-selected layer (L8) and best test layer on main EER (L10). Triangles: development EER above the axis range.}
\label{fig:layers}
\end{figure}

\subsection{XEUS on clean audio}
\label{ssec:profile}

Fig.~\ref{fig:layers} shows the clean results. Main EER falls from 16.9\% at L1 to its lowest values in L9--L12, then rises towards the top. The last layer is worse than L10 on all four seeds, by 1.99\,pp on main and 4.80\,pp on A14. Only the A14 gap survives singer resampling. Four seeds cannot separate the layers within L9--L12: L10 $-$ L11 is $-0.57 \pm 0.77$\,pp, and each seed has a different winner (L10, L12, L11, L9). The learned mix puts most of its weight on L6--L12 and almost none on L19 (0.014). It is worse than L10 (8.84\% vs.\ 7.50\%; 3 of 4 seeds) and better than the development-selected L8 on main (by 0.99\,pp), but worse than L8 on A14 (by 1.57\,pp) and under noise. Adding CQCC fusion gives the same profile and no consistent gain (fusion L10: 7.42\%).

\subsection{Choosing the layer on the development set}
\label{ssec:dev}

Development EER barely separates the middle layers. For XEUS it spans only 1.23--1.65\% across L3--L12, while main test EER spans 7.50--15.10\% over the same layers. The development and test rankings are weakly correlated over all 19 layers (Spearman $\rho = 0.29$, $p = 0.24$; 0.04--0.28 per seed) and not at all within L8--L13 ($\rho = -0.03$). With four seeds, the rule picks L8, which is worse than L10 on all four seeds, by 2.32\,pp on average; this gap survives singer resampling. With three seeds, the rule picks L11. Across all 15 subsets of the four seeds, it picks L8 ten times, L11 three times and L9 twice, usually by a tiny development-EER margin (median 0.056\,pp), and never picks the best test layer (mean loss 1.76\,pp). The learned mix has the lowest development EER of all heads (0.92\%) but ranks seventh of 20 on main test EER.
XLS-R shows the same problem more strongly: the rule picks L4 on both seeds, which is 4.13\,pp worse on clean main EER than the best test layer (L9), and $\rho = 0.36$ ($p = 0.08$). On A14 the encoders differ. For XEUS, the development and A14 rankings agree ($\rho = 0.74$), and the development pick L8 is also the best clean A14 layer (Sec.~\ref{ssec:a14}); for XLS-R they do not ($\rho = 0.12$).

\begin{table*}[t]
\centering
\caption{XEUS-only heads under corruption, test EER in \%. $n$: seeds. Best layer: lowest seed-mean test EER, chosen in hindsight. Differences: mean of per-seed paired differences in pp; in parentheses, the number of seeds with the same sign as the mean. L8 is the four-seed development pick; on the three seeds of the $n{=}3$ conditions, the rule picks L11. $^{*}$95\% singer-bootstrap interval excludes zero (not computed for MP3 128\,kbit/s and Opus 32\,kbit/s). $^\dagger$Under reverberation, L4 and L8 cannot be separated on A14, and the fusion heads' best A14 layer is L9.}
\label{tab:grid}
\small
\begin{tabular}{lc|cccc|cc}
\toprule
 & & \multicolumn{4}{c|}{Main EER (A09--A13)} & \multicolumn{2}{c}{A14 EER (out-of-domain attack)} \\
Condition & $n$ & Best layer & L10 EER & L19 $-$ L10 & L8 $-$ L10 & Best layer & L8 $-$ L10 \\
\midrule
Clean        & 4 & L10 & \phantom{0}7.50 & $+1.99$\phantom{$^{*}$} (4/4) & $+2.32^{*}$ (4/4) & L8 & $-0.54$\phantom{$^{*}$} (3/4) \\
Noise 20\,dB & 3 & L10 & 15.66 & $+7.34^{*}$ (3/3) & $+2.26^{*}$ (3/3) & L8 & $-2.18^{*}$ (3/3) \\
Noise 10\,dB & 3 & L10 & 22.64 & $+6.63^{*}$ (3/3) & $+1.40$\phantom{$^{*}$} (3/3) & L8 & $-3.23^{*}$ (3/3) \\
Noise 5\,dB  & 4 & L10 & 26.99 & $+5.05^{*}$ (4/4) & $+0.31$\phantom{$^{*}$} (3/4) & L8 & $-3.54^{*}$ (4/4) \\
Noise 0\,dB  & 4 & L9  & 31.57 & $+4.20$\phantom{$^{*}$} (4/4) & $+0.25$\phantom{$^{*}$} (3/4) & L8 & $-3.33^{*}$ (4/4) \\
Reverb (real RIR) & 4 & L12 & 19.46 & $-0.18$\phantom{$^{*}$} (2/4) & $+1.12$\phantom{$^{*}$} (3/4) & L4$^\dagger$ & $-3.12^{*}$ (4/4) \\
MP3 128\,kbit/s  & 3 & L10 & \phantom{0}7.62 & $+2.67$\phantom{$^{*}$} (3/3) & $+3.02$\phantom{$^{*}$} (3/3) & L8 & $-0.22$\phantom{$^{*}$} (2/3) \\
MP3 32\,kbit/s   & 3 & L10 & 10.64 & $+2.98^{*}$ (3/3) & $+3.14^{*}$ (3/3) & L11 & $+0.04$\phantom{$^{*}$} (1/3) \\
Opus 32\,kbit/s  & 3 & L12 & \phantom{0}9.12 & $+2.14$\phantom{$^{*}$} (3/3) & $+2.63$\phantom{$^{*}$} (3/3) & L8 & $-0.35$\phantom{$^{*}$} (2/3) \\
Opus 12\,kbit/s  & 3 & L11 & 13.69 & $+1.81$\phantom{$^{*}$} (3/3) & $+3.15^{*}$ (3/3) & L11 & $+0.36$\phantom{$^{*}$} (2/3) \\
\bottomrule
\end{tabular}
\end{table*}

\subsection{XEUS under corruption}
\label{ssec:corruption}

\begin{figure}[t]
\centering
\includegraphics[width=\columnwidth]{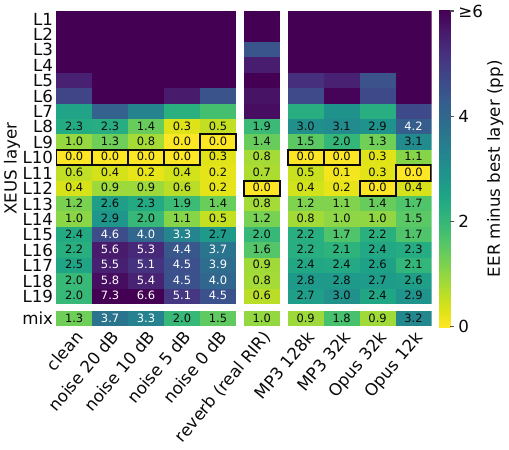}
\caption{XEUS-only heads: main EER of each layer minus that of the best layer in the same condition, for all ten test conditions (seed means; colour capped at 6\,pp, values shown for L8--L19 and the mix). Boxed: best layer per condition.}
\label{fig:grid}
\end{figure}

Table~\ref{tab:grid} summarizes the XEUS results under corruption, and Fig.~\ref{fig:grid} shows each layer's EER minus that of the best layer in each condition. On clean audio and under all noise and codec conditions, the best layer stays in L9--L12, as do 28 of the 30 per-seed best layers. Noise hurts every layer (L10 rises from 7.50\% to 15.66\% at 20\,dB) but hurts the upper layers most: L19 $-$ L10 grows from 1.99\,pp on clean audio to 4.2--7.3\,pp, with L10 better on every seed. MP3 at 128\,kbit/s barely changes the profile, and Opus at 12\,kbit/s hurts more than MP3 at 32\,kbit/s.

Reverberation behaves differently. All of L10--L19 reach 18.7--20.6\% main EER, L19 $-$ L10 is $-0.18 \pm 1.44$\,pp (L19 better on 2 of 4 seeds), and the per-seed best layers are L19, L12, L13 and L12. The upper layers thus catch up with the middle ones, rather than the best layer moving to a new depth. L1--L7 stay clearly worse (23--29\%). The learned mix is worse than L10 on every seed under noise (by 1.2--3.7\,pp) and under the two low-bitrate codecs (by 1.8 and 2.1\,pp).

\subsection{XEUS on the out-of-domain attack (exploratory)}
\label{ssec:a14}

A14 behaves differently. On clean audio its best XEUS layer is L8 (13.52\%), but its lead over L10 (0.54\,pp, 3 of 4 seeds) is within seed noise. Under noise, L8 stays best on A14 and its lead grows to 2.2--3.5\,pp, on every seed at every SNR, and the gap survives singer resampling at every SNR and under reverberation (Table~\ref{tab:grid}). Choosing the A14 layer on either half of the test clips also gives L8, which then wins on the other half, and the held-out fourth seed agrees ($-4.05$ and $-3.26$\,pp at 5 and 0\,dB). On main EER, by contrast, L8 stays 0.3--2.3\,pp worse than L10 under noise.

Under reverberation, the best A14 layer moves earlier, into L4--L8 (L4: 27.42\%, L8: 28.06\%, L10: 31.19\%). Four seeds cannot separate L4 from L8 ($t = -0.66$), and the shift appears only for the XEUS-only heads; with CQCC fusion, the best A14 layer under reverberation is L9. Under the two low-bitrate codecs, the best A14 layer is L11, inside the main-EER band. Because we chose these comparisons after seeing three of the four seeds, $p$-values would overstate the evidence even after a Bonferroni correction for the 20 comparisons. We rely instead on effect sizes, seed agreement, split-half stability and the held-out seed, which was scored on the same test clips.

\begin{figure}[t]
\centering
\includegraphics[width=\columnwidth]{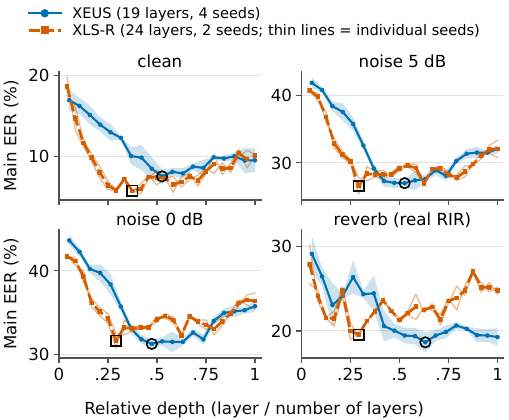}
\caption{Main test EER against relative depth (layer index / number of layers) on clean audio, 5 and 0\,dB noise and reverberation. XEUS: seed mean $\pm$ sample std (4 seeds); XLS-R: seed mean (bold) and individual seeds (thin, 2 seeds). Outlined markers: best seed-mean layer in each condition.}
\label{fig:depth}
\end{figure}

\subsection{A second encoder: XLS-R}
\label{ssec:xlsr}

Fig.~\ref{fig:depth} compares the two encoders by relative depth. Three XEUS results hold for XLS-R. Its last layer is 4.4--5.5\,pp worse than its best layer in every condition, on both seeds, and the gap survives singer resampling. The learned mix does not beat the best single layer. On clean audio the best layers sit in the lower middle (L9, relative depth 0.38, 5.71\%; the per-seed best are L10 and L14, at depths 0.42 and 0.58). Two results differ. Under noise, XLS-R has no stable best layer: the two seeds pick L7 and L15, with nearly flat EER between them. Under reverberation, XEUS's upper layers catch up (last minus best: 0.62\,pp), whereas XLS-R's best layers move down to L6--L7 on both seeds (L7: 19.57\%) and its upper layers get worse: relative to L7, L10 by 3.7--4.4\,pp and L24 by 5.0--5.4\,pp. Finally, XLS-R's A14 EER is nearly flat across depth on clean audio (L5--L21: 11.5--12.5\%), so the A14 result of Sec.~\ref{ssec:a14} is specific to XEUS.

\section{Discussion and Limitations}
\label{sec:discussion}

In both encoders, a middle layer is a better clean-audio readout than the last one, but where the best layer sits, and how it moves under corruption, depends on the encoder. With this recipe, the learned mix does not match the best single layer on main EER. This does not show that a single layer is always better: the XEUS mix does beat the development-selected layer on clean main EER, though not on A14 or under noise.

Choosing the layer on the development set costs 2.3\,pp (XEUS) and 4.1\,pp (XLS-R) of clean main EER. This does not mean development data is useless: for XEUS, the development pick L8 is also the best clean A14 layer. The development set may fail because its error rates are too low to rank layers, because its attacks differ from the test attacks, or both; our experiments cannot tell these apart. Validating on held-out attack families or on corrupted audio may help, but we have not tested this.

For XEUS, L9--L12 work well on A09--A13 under noise and coding; under reverberation, L10--L19 are about equally good, whereas XLS-R prefers shallower layers. This fits XEUS's dereverberation pretraining~\cite{xeus2024}, but the encoders also differ in architecture, training data and size, so we cannot attribute the difference to that objective.

The study covers one singing-voice dataset, two frozen encoders and one head recipe. XLS-R has only two seeds, four test conditions and no CQCC fusion, so we describe the two encoders rather than rank them. Main-EER spoofs come from 11 singers, so several main-EER differences, including the clean XEUS L19 $-$ L10 gap, do not survive singer resampling. We use one noise corpus, one set of real room impulse responses and two codecs, with no augmentation in training. XEUS has 3--4 seeds per condition, and seed-to-seed variation mixes initialization and data order. The A14 analysis is post hoc, and its extra seed was scored on the same test clips; attack and source effects are not separated. Music-pretrained encoders and fine-tuned encoders remain untested.

\section{Conclusion}
\label{sec:conclusion}

On CtrSVDD, the middle layers of two frozen speech encoders detect singing-voice deepfakes better than their last layers on clean audio, and a learned layer mix does not beat the best single layer. Choosing the layer on the development set misses the best test layer in both encoders. How the best layer responds to corruption depends on the encoder, and for XEUS the out-of-domain attack favours a different layer than the other attacks. Layer-wise results should therefore be reported across conditions and attack types, rather than assuming that the last layer or the layer with the lowest development error will transfer.


\section{Compliance with Ethical Standards}
This study uses publicly available data under their respective licences: CtrSVDD (CC~BY-NC-ND~4.0), whose bona fide source corpora carry their own terms; MUSAN (CC~BY~4.0); and RIRS\_NOISES (Apache~2.0), whose real impulse responses come from RWCP, AIR and REVERB~2014 under their own terms. No audio is redistributed. No new human participants were recruited or recorded for this study.

\section{Acknowledgments}
This work was supported in part by computational resources
provided by the NSF ACCESS program under allocation
CIS261055. We thank James O'Brien of Sound Ethics for his
mentorship.
Claude (Anthropic) and ChatGPT (OpenAI) supported manuscript clarity and revision. The authors also used such tools to audit code and write plotting code for Figures 1--3. After using these tools, the authors reviewed and edited the content as needed and take full responsibility for the publication's content.

\end{document}